# Enhancement of SHG efficiency by urea doping in ZTS single crystals and its correlation with crystalline perfection revealed by Kurtz powder and high-resolution X-ray diffraction methods

G. Bhagavannarayana and S. K. Kushwaha

*Materials Characterization Division, National Physical Laboratory, New Delhi-110 012, India*

**Synopsis**  Urea doping lead to enhance the SHG efficiency in ZTS single crystals with a critical dependence on its concentration. The interesting correlation between crystalline perfection and SHG efficiency has been revealed by using high-resolution X-ray diffractometry and Kurtz powder techniques.

**Abstract**  Enhancement of second harmonic generation (SHG) efficiency and the correlation between crystalline perfection and SHG with urea doping on tristhioureazinc(II) sulphate (ZTS) single crystals have been investigated. ZTS is a potential semiorganic nonlinear optical (NLO) material. Pure and urea doped single crystals of ZTS have been successfully grown by slow evaporation solution technique (SEST). Presence of dopants has been confirmed and analyzed by Fourier transform infrared (FTIR) spectrometer. The influence of urea doping at different concentrations on the crystalline perfection has been thoroughly assessed by high resolution X-ray diffractometry (HRXRD). HRXRD studies revealed that the crystals could accomodate urea in ZTS up to some critical concentration without any deterioration in the crystalline perfection. Above this concentration, very low angle structural grain boundaries were developed and it seems, the excess urea above the critical concentration was segregated along the grain boundaries. At very high doping concentrations, the crystals were found to contain mosaic blocks. The SHG effeiciency has been studied by using Kurtz powder technique. The relative SHG efficiency of the crystals was found to be increased substantially with the increase of urea concentration.  The correlation found between the crystalline perfection and SHG efficiency was discussed.

## 1. Introduction

Due to high-speed and ease of production of photons (light), the area of photonics has become an active field of research in view of modern society's demand for improved telecommunications, data retrieving, storage, processing and transmission. The design of devices that utilize photons instead of electrons in the transmission of information has created a need for new materials with unique optical properties (Williams & Angew, 1984). Hence, it will be useful to synthesize new NLO materials and study their structural, physical, thermal and optical properties. It is also equally important to enhance the NLO properties of the known materials by either the incorporation of functional groups (Sweta *et al.*, 2007; Ushasree *et al.*, 1999) or dopants (Bhagavannarayana *et al.*, 2008), for tailor made applications. ZTS is a well characterized (Andreettie *et al.*, 1968) material of noncentrosymmetric orthorhombic crystal system with lattice parameters a = 11.126 Å, b = 7.773 Å and c = 15.491 Å and space group *Pca*$2_1$ (point group *mm*2). It exhibits a low angular sensitivity, and hence useful for second harmonic generation (SHG) having SHG efficiency of 1.2 times than that of potassium dihydrogen phosphate (Marcy *et al.*, 1992). High laser damage threshold and wide optical transparency were found (Venkataramanan *et al.*, 1995). Thermal, elastic and electro-optic (Kerkoc *et al.*, 1996; Alex & Philip, 2001; Sastry, 1999) properties were also been reported. Better optical properties were found by mixing phosphate in ZTS crystals (Ushasree *et al.*, 1999). Our recent studies on ZTS (Bhagavannarayana *et al.*, 2008; Bhagavannarayana *et al.*, 2006) in the presence of some inorganic/organic dopants elucidated the enhancement of crystalline perfection which in turn leads to the improvement in the SHG efficiency. Due to such inherent advantages and continuous miniaturization of the modern devices, crystalline perfection becomes more and more stringent in the device technology of modern science. To realize the full efficiency of the properties, the crystals should be free from defects (Bhagavannarayana, *et al.*, 2005). In the present investigation, effect of urea (a NLO material) doping in ZTS crystals has been studied. The presence of urea in doped ZTS crystals has been confirmed by FTIR. The crystalline perfection of undoped and urea doped crystals at different concentrations has been evaluated by HRXRD. The effect of urea doping on SHG efficiency was studied by Kurtz powder technique. Correlation between the crystalline perfection and SHG efficiency was well discussed.

## 2. Experimental

### 2.1. Crystal Growth

The title compound ZTS was synthesized from initial materials zincsulphateheptahydrate and thiourea by taking in 1:3 stoichiometric ratio in deionized water according to the following chemical reaction (Arunmozhi *et al.*, 2004; Bhagavannarayana *et al.*, 2006). The low temperature (333 K) of the solution was maintained to avoid decomposition of ZTS.

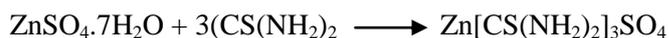

$$ZnSO_4 \cdot 7H_2O + 3(CS(NH_2)_2) \longrightarrow Zn[CS(NH_2)_2]_3 SO_4$$

The resultant product was recrystallized to purify it and the saturated mother solution was prepared at 300 K with slight acidic condition (pH~5.9) (Bhagavannarayana *et al.*, 2008; Ushasree *et al.*, 1999). Urea with different concentrations (0.1 to 12 mol%) has been added separately. Pure and doped ZTS solutions were continuously stirred for around 12 hours for homogeneous mixing of dopant. The beakers which contain these saturated solutions were mounted in a constant temperature bath with the constant growth parameters at 300 K for slow evaporation. The growth conditions were closely monitored. Within a span of 20 days, good quality pure and doped single crystals were harvested. The structure of the grown crystals was confirmed by powder XRD analysis.

## 2.2. FTIR Spectroscopy

To confirm the presence of functional groups of ZTS and also to study the incorporation of urea in the doped crystals, the specimens with different concentrations of urea were characterized by FTIR spectrometer (Nicolet-5700) at room temperature in the wavenumber range of 500-3500 cm$^{-1}$. The FTIR spectra for pure and urea doped ZTS are shown in Figure 1.

## 2.3. High-resolution Multicrystal X-ray diffractometry

To reveal the crystalline perfection of the grown crystals and to study the effect of dopants added in the saturated aqueous solution during the growth process, a multicrystal X-ray diffractometer (MCD) developed at NPL (Lal & Bhagavannarayana, 1989) has been used to record high-resolution diffraction curves (DCs). The schematic of this system is shown in Figure 2 (a). In this system, a fine focus (0.4 × 8 mm$^2$; 2 kW Mo) X-ray source energized by a well-stabilized Philips X-ray generator (PW 1743) was employed. The well-collimated and monochromated MoK$\alpha_1$ beam obtained from the three monochromator Si crystals set in dispersive (+,-,-) configuration has been used as the exploring X-ray beam. The fineness and the uniformity throughout its length may be seen in the inset of Figure 2(a). The dispersive nature of this configuration may be understood from Figure 2(b) (Bhagavannarayana, 1994). Curves A and B in Figure 2 (b) show the DCs of the third monochromator recorded in non-dispersive (+,-,+) and dispersive (+,-,-) modes respectively. One can see that in the dispersive configuration, the slightly divergent parts of the beam disperse away towards the tails or wings of the DC. It is interesting to compare the FWHM and peak intensity of these curves. In curve B, the FWHM is increased three times whereas the peak intensity is decreased around one third in comparison to that of curve A showing that the areas under both the curves (integrated intensities) are same which in turn reveals no loss of beam, but the slightly divergent portions of the beam dispersed away from the exact peak position. This arrangement improves the spectral purity ($\Delta\lambda/\lambda << 10^{-5}$) of the MoK$\alpha_1$ beam. The divergence of the exploring beam in the horizontal plane (plane of diffraction) was estimated to be << 3 arc sec. The specimen crystal is aligned in the (+,-,-,+) configuration. Due to dispersive configuration, though the lattice constant of the monochromator crystal(s) and the specimen are different, the unwanted dispersion broadening in the diffraction curve of the specimen crystal is insignificant. The specimen can be rotated about a vertical axis, which is perpendicular to the plane of diffraction, with minimum angular interval of 0.4 arc sec. The diffracted intensity is measured by using a scintillation counter. The rocking or diffraction curves (DCs) were recorded by changing the glancing angle (angle between the incident X-ray beam and the surface of the specimen) around the Bragg diffraction peak position $\theta_B$ (taken zero as reference point) starting from a suitable arbitrary glancing angle ($\theta$). The detector was kept at the same angular position $2\theta_B$ with wide opening for its slit, the so-called ω scan. In the present study, the X-ray power, size of the beam and configuration of the diffractometer were kept same for all the specimens throughout the experiments. Before recording the diffraction curve, to remove the non-crystallized solute atoms remained on the surface of the crystal and also to ensure the surface planarity, the specimens were first lapped and chemically etched in a non preferential etchant of water and acetone mixture in 1:2 volume ratio.

**2.4. Kurtz powder SHG measurements**

SHG test on the crystals was performed by Kurtz powder method (Kurtz & Perry, 1968). An Nd: YAG laser with a modulated radiation of 1064 nm was used with 8 ns pulse width and 10 Hz repetition rate as the optical source and directed on the powdered sample through a filter. The grown crystals were ground to a uniform particle size of 125 – 150 μm which is much more than that of the coherence length of laser beam, and then packed in a microcapillary of uniform bore and exposed to laser radiations. Second harmonic radiation generated by the randomly oriented microcrystals was focused by a lens and detected by a photomultiplier tube after filtration of the incident or fundamental radiation of 1064 nm. The doubling of frequency was confirmed by the green colour of the output radiation whose characteristic wavelength is 532 nm. The experimental settings were kept same for all the specimens to analyze the relative influence of doped urea in the crystal on SHG efficiency.

**3. Results and discussion**

**3.1. FTIR analysis**

The concentration of entrapped dopants in the crystal most likely not to be the same as in the solution due to the fact that while growing the crystal, it has a tendency to reject the foreign atoms or molecules to enter into the crystal lattice unless until they are chemically very favourable (like valancy, size, chemical/hydrogen bonding) and hence the real concentration of the dopants accommodated or entrapped in the crystal lattice may be much lesser but expected to be proportional to the prevailing concentration in the solution. Though it is in principle possible to determine the true concentration in the crystal by sensitive characterization tools like atomic absorption spectroscopy, inductive coupled plasma, X-ray fluorescence spectroscopy etc., in the present case as the molecules of thiourea in the ZTS crystal and the molecules of urea (dopant) contain mostly the same atoms (C) or groups ($NH_2$) and hence these techniques could not yield the correct concentration of dopants. But using the stretched C=O bond in urea, one can not only confirm the presence of urea but also one can get an idea of relative quantity of incorporated urea in the crystal by the relative prominence of the absorption band in FTIR spectra. Figure 1 shows the FTIR spectra of pure and urea doped ZTS specimens with different concentrations ranging from 0.1 to 7.5 mol%. The peaks at 1628, 1502, 1404 and 714 cm$^{-1}$ indicate $NH_2$ bending, N–C–N stretching, C=S asymmetric stretching and C=S symmetric stretching bonds respectively as expected in pure ZTS crystals (Meenakshisundaram *et al.*,

2006). The observed absorption peaks at 1736 and 1210 cm$^{-1}$ (as indicated by the dotted lines) indicate the stretched C=O and C–O bonds (Szetsen *et al.,* 1999; Wu *et al.,* 2003) respectively. The absence of shift of C=O absorption band indicates the incorporation of urea in the interstitial position instead of substitutional position. For pure and urea doped (0.1 and 1.0 mol%) crystals, the peaks are not well resolved. But above these concentrations, one can see the well resolved peaks with increasing prominence of these absorption bands due to increase in urea concentration. These features confirm the incorporation of urea in the crystalline matrix. The gradual increase in the prominence of these bands confirms the fact that due to increase in the concentration of urea in the solution, incorporation of urea in the crystal also proportionally increased. The occurrence of absorption band due to C-O indicates the presence of hydrogen bonds due to the presence of $NH_2$ groups of thiourea in ZTS matrix (Abhay Shukla *et al.*, 2001; Yuji Kohno *et al.*; 2003, Luo Ning e*t al.*, 1997). There are good number of examples in the literature (Dongfeng Xue & Siyuan Zhang, 1970; Midori Kato e*t al.*,1997; Xue Dongfeng & Zhang Siyuan, 1996) which confirm that hydrogen bonding becomes the cause for the NLO nature of the crystals or helpful to enhance it. The same result of enhancement of SHG has been observed experimentally in our present investigation as described in the forthcoming section and hence confirms the hydrogen bonding in urea doped ZTS crystals. These hydrogen bonds also help the entrapment of urea interstitially in the crystal and thereby help in enhancing SHG efficiency (Midori Kato *et al.*, 1997) which is otherwise not possible as urea cannot occupy easily the substitutional position of thiourea. The investigations by powder XRD, HRXRD and SHG also confirm the same as described in the forthcoming sections.

### 3.2. High-resolution XRD analysis

Before proceeding for the HRXRD studies, powder XRD analysis for undoped and urea doped specimens was carried out. The structure and the lattice parameters of both undoped and urea doped crystals were found to be the same as reported (Andreettie *et al.,* 1968). Except the minor variations in the peak intensities of different spectral lines due to strains, neither additional phases nor significant variation in lattice parameters were found due to urea doping. In order to analyze the effect of dopants on the crystalline perfection, high-resolution X-ray diffraction curves (DCs) were recorded as explained in § 2.3. As shall be seen in the forthcoming analysis, depending upon the nature of DCs which in turn depend on the degree of concentration of dopants, the specimens are categorized in the following three groups: (i) Undoped specimen, (ii) Specimens doped with concentrations upto 2.5 mol% and (iii) Specimens doped with concentrations between 2.5 to 12 mol%.

### 3.2.1. Undoped specimen

Figure 3(a) shows the DC for the undoped ZTS crystal recorded for (200) diffracting planes using MoK$\alpha_1$ radiation in symmetrical Bragg geometry. The diffracted intensity of this curve and the other DCs to be described in the forthcoming sub sections are arbitrary, but the magnitude is relative. The experimental conditions like power and size of the X-ray beam are same and no normalization to either the peak area or the peak intensity is made. The range of the glancing angle for the DCs is so chosen to cover the meaningful scattered intensity on the both sides of the peak. The unit of glancing angle is in arc s. It may be mentioned here that to assess the crystalline perfection one can choose any convenient set of planes which in turn covers the entire volume of the crystal. ZTS specimens grow with major surfaces along [100] and (200) planes give the diffraction. The diffraction curve of Figure 3(a) is quite sharp having FWHM of 5 arc sec with a good symmetry with respect to the exact Bragg diffraction peak position (set as zero). Such a sharp curve is expected for a nearly perfect single crystal according to the plane wave dynamical theory of X-ray diffraction (Batterman & Cole, 1964). The sharp and single peak indicates that the specimen does not contain any internal structural grain boundaries (Bhagavannarayana *et al.,* 2005). The scattered intensity along the wings/tails on both sides of the exact Bragg peak (zero glancing angle) of DC is quite low, showing that the crystal does not contain any significant density of dislocations and point defects and their clusters (Lal & Bhagavannarayana, 1989). These features reveal that the quality of the pure ZTS is quite high. The quality is further tested by measuring the radius of curvature as given below.

To see the flatness of the crystallographic planes of the grown crystals, radius of curvature has been determined by recording the change in the diffraction peak position for the desired planes with respect to the linear position of the specimen as the specimen is traversed across the incident/exploring beam (Lal *et al.,* 1990). Figure 3(b) shows such a plot for same specimen as that of Figure 3(a). It may be mentioned here that the initial Bragg peak position which was set at 100 arc s is arbitrary and the slope does not depend on this value. The radius of curvature for (100) crystallographic planes of the specimen obtained by the reciprocal of slope of this plot is 1.51 km. This value is quite high which is expected for a good quality flat crystal (Sharma *et al.,* 2006). It may be mentioned here, the quantitative measurement of such flat crystals is not possible with the desired accuracy when the FWHM values are not low. In such a case the uncertainty in the location of peak position is high to determine quantitatively the value of radius of curvature for such flat crystals.

### 3.2.2. Specimens doped with concentrations up to 2.5 mol%

To analyze the effect of dopants in this range, three specimens with concentrations 0.1, 1.0 and 2.5 mol% were studied. The DCs of these specimens are shown in Figure 4. As mentioned above, the relative diffracted X-ray intensity for all the samples is same as the experimental conditions like power and size of the X-ray beam are same and no normalization to either the peak area or the peak intensity is made. As seen in Figure 4, all the three diffraction curves are having single peaks as in Figure 3 (a), confirming the fact that these doped specimens also do not have any structural grain boundary. However, the FWHM gradually increases as the urea (dopant) concentration increases. As seen in Figure 4, FWHM values for the specimens with concentrations 0.1, 1.0 and 2.5 mol% respectively are 13, 18 and 21 arc sec. These are quite high in comparison to 5 arc sec belongs to the undoped specimen showing that the doping has a significant influence on the value of FWHM. On careful observation, one can also see that the intensity increases sharply as the glancing angle approaches the peak position as expected for a perfect crystal. But at higher glancing angles (away from the Bragg peak position), the scattered intensity falls down slowly. For the sake of convenient comparison, all these three DCs in Figure 4 along with the DC of undoped specimen given in Figure 3 (a) are combindly drawn in Figure 5 with a dotted straight line at the exact Bragg peak position. A common range for the glancing angle from -100 to 100 is chosen for all the curves so as to see the asymmetry of the curves with respect to the peak position. From this figure one can clearly see that the intensity along the wings/tails of the DCs gradually increased as the dopant concentration increased. The increase in FWHM without having any additional peaks indicates the incorporation of dopants in the crystalline matrix of ZTS. The gradual increase of FWHM and scattered intensity along the wings of the DCs as a function of prevailing concentration of urea in the solution during the growth process indicate that the actual amount of urea entrapped (or doped) in the crystal is proportional to the concentration of urea present in the solution which is also in tune with the FTIR results. It is interesting to observe few features of these curves: (i) the peak intensity, (ii) the area under the curve also known as integrated intensity ($\rho$) and (iii) the asymmetry of the curves. The peak intensity of these curves rapidly decreases to 1 mol% and is saturated from 1 to 2.5 mol% whereas the FWHM value rapidly increases to 1mol% and is almost saturated from 1 to 2.5 mol% doping. However, $\rho$ remains almost the same. It may be mentioned here that when the crystal is imperfect with mosaic blocks, $\rho$ is expected to be very high (James, 1950) as it is proportional to $F^2_{hkl}$ which is otherwise proportional to $F_{hkl}$ ($F_{hkl}$ being the structure factor) for a perfect crystal. The constancy of $\rho$ with increase in dopant concentration indicates that the dopants are not agglomerated into mosaic blocks but statistically distributed in the crystal lattice. But the value of $\rho$ for sample with 2.5 mol% is slightly higher than that of other specimens which indicates that this concentration is also slightly higher than that of the critical concentration up to which the dopants can stay in the crystal in isolated form without agglomeration. In DCs of doped specimens, for a particular angular deviation ($\Delta\theta$) of glancing angle with respect to the peak position, the scattered intensity is relatively more in the positive direction in comparison to that of the negative direction. This feature or asymmetry in the scattered intensity clearly indicates that the dopants predominantly occupy the interstitial positions in the lattice and elucidates the ability of accommodation of dopants in the crystalline matrix of the ZTS crystal. This can be well understood by the fact that due to incorporation of dopants in the interstitial positions, the lattice around the dopants compresses and the lattice parameter d (interplanar spacing) decreases and leads to give more scattered (also known as diffuse X-ray scattering) intensity at slightly higher Bragg angles ($\theta_B$) as d and sin $\theta_B$ are inversely proportional to each other in the Bragg equation ($2d \sin \theta_B = n\lambda$; n and $\lambda$ being the order of reflection and wavelength respectively which are fixed). It may be mentioned here that the variation in lattice parameter is only confined very close to the defect core which gives only the scattered intensity close to the Bragg peak. Long range order cannot be expected and hence change in the lattice parameters also cannot be expected as we could not found any change in powder XRD.

Entrapment of small amounts of dopants though they cannot substitute any host atom or molecule is possible due to their presence in the solution at large quantities. For molecules like urea in the host crystal like ZTS, the possible hydrogen bonds also help for their entrapment in small quantities. Entrapment in the interstitial positions is elucidated by the observed pronounced scattering on the higher diffraction angles with respect to the Bragg peak position. If urea would have taken the substitutional position of thiourea, lattice around the defect core (i.e. urea) might have widened (as S atoms in thiourea are larger than O atoms in urea) and experimentally one would get pronounced scattering on the lower diffraction angles. But experimentally, the other way is found and hence the occupation of urea in the interstitial position of the lattice with an associated compressive or compositional strain is a compatible conclusion of these findings. The correlation between dopant concentration with FWHM, ρ and asymmetry of the diffraction curve at lower amount of urea doping is indeed possible due to the high-resolution of the multicrystal X-ray diffractometer used in the present investigations. Otherwise one cannot distinguish such small variations in the FWHM particularly when the concentration of dopants or defects is very low. Here the used wavelength of the X-ray probe is 0.709261 Å (MoK$\alpha_1$) and hence the present studies pertaining to even a dense structure of ZTS, minute details of structural variation aroused due to dopants could be observed. As mentioned above, because of the entrapment of dopants (urea) in the interstitial positions of the crystal, the local region i.e. the region around the defect core undergoes compressive strain leading to reduction in the d spacing. Because of this, one expects scattering from the local Bragg diffraction from these defect core regions. Indeed, the d spacing of the whole crystal is not expected to change due to short range order of such strain. Therefore, omega scan used in the present investigation is good enough to collect all the scattered or local Bragg intensities due to such strained regions and can be attributed to the local compressive/compositional strain by the entrapped urea. Some more useful details may be found in our recent article pertaining to the studies on dopants in ADP crystals (Bhagavannarayana *et al.,* 2008). The effect of $Cr^{3+}$, $Fe^{3+}$ and $Al^{3+}$ on ADP crystals has been studied (Comer, 1959; Mullin *et al.,* 1970; Davey & Mullin, 1976) and it is known from Mossbauer studies (Fontcuberta *et al.,* 1978) that incorporation takes place at interstitial lattice sites.

### 3.2.3. Specimens doped with concentrations between 5 to 12 mol%

In this range of dopant concentration, the experimentally observed DCs contain additional peak(s). The curves (a), (b) and (c) in Figure 6 show respectively the DCs of three typical specimens whose urea concentration is 5, 7.5 and 12 mol%. These DCs have quite different features than that of DCs in Figures 3(a) and 4. In addition to the main peak at zero position, these curves contain additional peak(s). The solid line in the curves (a) and (b) which is well fitted with the experimental points is obtained by the Lorentzian fit. The additional peaks at 24 and 36 arc sec away from the main peak respectively in curves (a) and (b) are due to internal structural very low angle (≤ 1 arc min) grain boundaries (Bhagavannarayana *et al.,* 2005). The tilt angle i.e. the misorientation angle of the boundary with respect to the main crystalline region for these very low angle boundaries are 24 and 36 arc sec. Though these tilt angles (which are in arc secs) are very small, they indicate that the heavy compressive stress due to urea dopants at the high concentrations like 5 and 7.5 mol% lead to grain boundaries in the crystal. To rule out the possible epitaxial growth in SEST grown crystals (Bhagavannarayana *et al.,* 2006), few micron surface layer of the crystal were ground and lapped and recorded the DC. But still additional peak persists with the same tilt angle. Since the tilt angle is in the order of few arc seconds, one cannot attribute these grain boundaries as twins. For further confirmation, section topographs were recorded separately at both the peaks of the DCs. As a typical example, for 5 mol% specimen, the section topographs were recorded separately at both the peaks of Figure 6. The topographs indicted by I and II in Figure 7 are respectively correspond to the very low grain boundary and the main crystal region. The grain type of dark background in this figure is due to poor resolution of photo occurred mainly because of the huge enlargement of the photograph. The size of the exploring X-ray beam on the X-ray film is 5 mm x 0.2 mm as indicated in the figure. As seen in the topographs, the intensity is not uniform along the length. In the left hand side topograph belongs to the sharp peak at 24 arc sec, one can see good intensity on top portion. On the other hand, the bottom portion in the right hand side topograph contains more intensity. These observations indicate that the top and bottom crystalline regions of the specimen are mis-oriented by 24 arc sec and confirm the fact that the additional peak is due to a very low angle structural tilt grain boundary. Similarly, the additional peak in curve (b) also depicts the very low angle boundary. The high values of FWHM for the main peaks of these two specimens (having 5 and 7.5 mol% urea concentrations which are respectively 55 and 110 arc sec) indicate that the quality of these regions is not up to the mark. The large values of FWHM of the main peaks of curves (a) and (b) do not rule out the absence of mosaic blocks, which are misoriented to each other by few arc sec to few tens of arc sec. The consisting observation regarding FWHM is that more the dopant concentration, more its value. In these curves, it is also interesting to note down the lower FWHM values of 10 and 22 arc sec for additional peaks. Such low values of FWHM indicate that during the growth process, the entrapped dopants in the crystalline matrix slowly moved towards the nearby boundary and segregated along them. The heavy compressive stress seems to be the driving force for the movement of the excess dopants by the process of guttering. Such type of segregation of dopants along the boundaries was well confirmed in our earlier studies by SIMS on BGO crystals wherein the Si impurities were found to segregate along the structural grain boundaries (Choubey *et al.,* 2002). But the crystalline regions on both sides of the boundary contain some amount of dopant (urea), which may be a critical concentration to accommodate in the crystal and is responsible for the observed enhancement of SHG compared to that of the undoped or doped crystals at low concentration as observed in the forthcoming section. The segregated urea however, cannot contribute anything for the enhancement as it does not exist in crystalline state though the urea itself is a good NLO material. But as mentioned above, the crystalline regions on both sides of the boundary which contains some entrapped urea in isolated form in the interstitial positions of the crystal contributes SHG of ZTS. However, such crystals with boundaries though they also show enhancement of SHG may not be of much use as the crystals needed for devices should be defect free for stability, reliability and full yield of SHG.

Curve (c) in Figure 6 is the DC recorded for a specimen with 12 mol% urea. In the angular range between 100 to 500 arc s, a mixture of unresolved low intensity peaks can be seen and reveals the fact that the specimen contains a good number of mosaic blocks, which might be formed due to release of heavy stress aroused in the crystal from heavy doping. However, as in curves (a) and (b), this curve also contains one sharp peak which seems to be due to the denuded crystalline region from excess urea. As explained above such crystals are not good for device fabrication as crystals are anisotropic in nature and give full yield of their output only when all parts of the crystalline regions are having the same crystallographic orientation. The HRXRD results confirm an important finding that urea can be entrapped in the ZTS crystals, but the amount is limited to a critical value and above which the crystals have a tendency to develop structural grain boundaries. The excess urea entrapped in the heavily doped crystals seems to be segregated along the boundaries by the process of guttering and as a result of it; some regions are denuded from the excess dopants.

**3.3. SHG efficiency**

As described in § 2.4, SHG test on the powder samples was performed by Kurtz powder SHG method with the Input radiation of 2.7 mJ/pulse. Output SHG intensities for pure and doped specimens give relative NLO efficiencies of the measured specimens. These values are given in **Table-1** along with the output values of urea and KDP. As seen in the table, SHG output enhances considerably with the urea doping which is one of the most important findings of the present investigation. It is worth to mention here about the possible correlation of SHG output on crystalline perfection. As seen in the table, three distinct specimens of ZTS were chosen for the SHG measurements. As found in the HRXRD studies both undoped and 2.5 mol% urea doped ZTS are having good crystalline perfection. Whereas 7.5 mole % urea doped specimen contain structural tilt grain boundaries which are detrimental to the NLO character as mentioned above. However, as seen in the table, urea doping enhances the SHG output irrespective of crystalline perfection. Our recent studies (Bhagavannarayana *et al.,* 2006) on ZTS and ADP crystals show a direct bearing of crystalline perfection on SHG efficiency. The controversy can be realized in the following way. In the present investigation, as observed in HRXRD studies, even in the heavily doped crystals, some portions (denuded regions from excess urea) of the crystal contains good crystalline perfection and contribute to the enhancement of SHG of the ZTS crystal. However, one should not ignore the crystalline perfection which deteriorates when the concentration is very high, due to formation of structural boundaries and leads to decrease in SHG efficiency as the total SHG yield from the different grains of the crystal with different orientation is expected to be less as SHG is anisotropic in single crystals.

## 4. Conclusions

FTIR studies confirm the incorporation of urea dopant in ZTS crystal by adding urea in the solution while crystal is growing through slow evaporation of the solution. These studies also indicate the presence of hydrogen bonds in the doped crystals. From the HRXRD studies, it is clearly demonstrated that the crystalline perfection strongly depends on the dopant concentration. Depending upon the size and nature of the dopants, there is a limit of dopant concentration below which the crystal can accommodate. Above that limit, the dopants lead to develop structural grain boundaries and segregate along the boundaries by the heavy compressive stress in the lattice developed by them. Urea doping leads to increase SHG efficiency of the ZTS crystals substantially. It was also concluded that when we use certain dopants to increase the SHG efficiency of the host crystal, one should also cautious about the crystalline perfection as it deteriorates considerably (by the formation of structural grain boundaries) at higher concentrations without much reduction in SHG efficiency particularly when the SHG output is measured by powder technique as observed in the present study. But in case of single crystals having structural boundaries, the total SHG output when measured in single crystal form certainly decrease as the SHG is a directional property.

**Acknowledgements**  The authors acknowledge Ex-director, NPL, Dr. Vikram Kumar for his continuous encouragement in pursuing the above studies. Kushwaha would like to acknowledge Council of Scientific and Industrial Research (CSIR) for providing the Senior Research Fellowship (SRF).

**Table 1** Relative second harmonic generation (SHG) output.

| Specimen | SHG output (mV) |
|---|---|
| Urea | 715 |
| KDP | 119 |
| ZTS undoped | 143 |
| ZTS doped with 2.5 mol% urea | 242 |
| ZTS doped with 7.5 mol% urea | 287 |

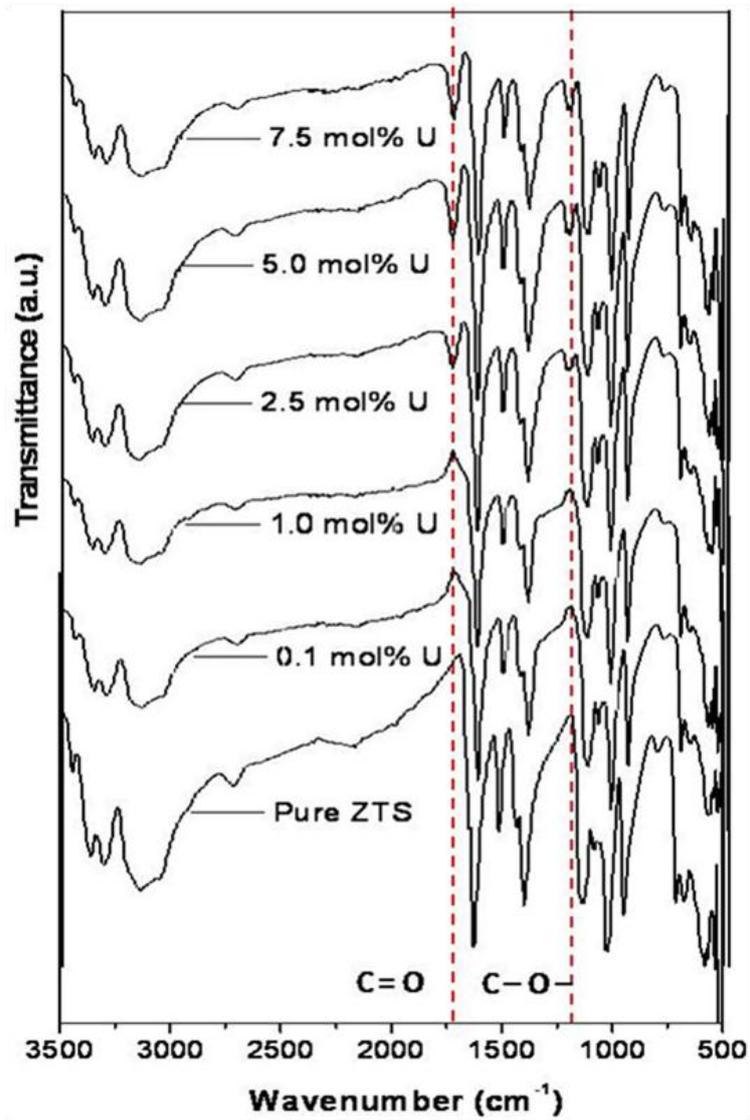

**Figure 1** FTIR spectra for pure and urea (U) doped ZTS specimen.

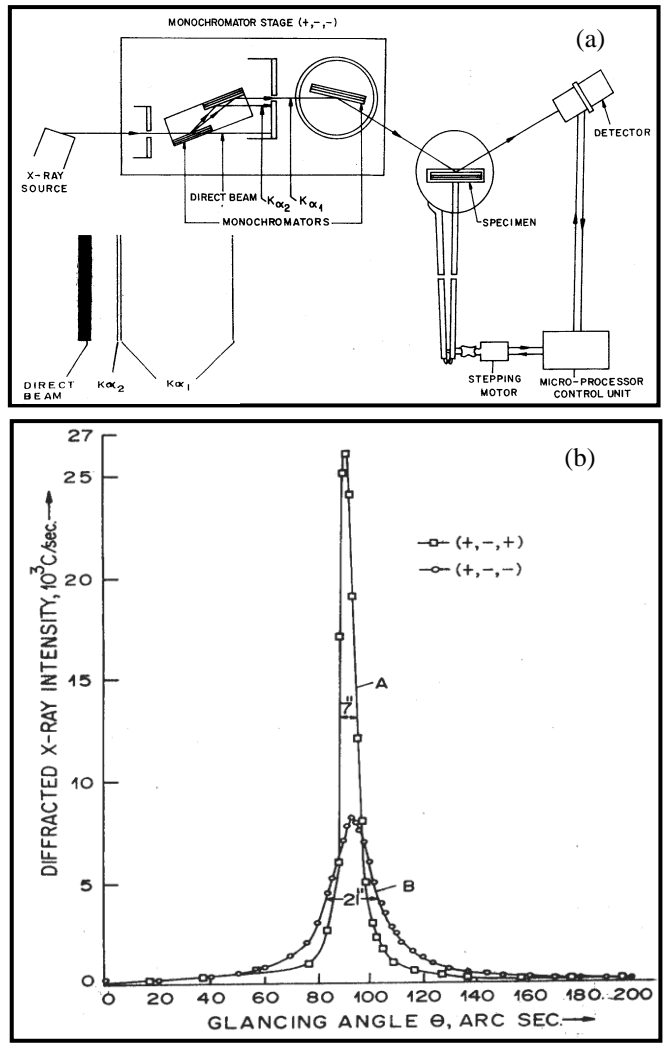

**Figure 2** (a) Schematic of the multicrystal X-ray diffractometer developed at NPL. Inset shows the MoKα doublet and the isolated Kα$_1$ beam and (b) DCs recorded in dispersive (+,-,-) and non-dispersive (+,-,+) settings.

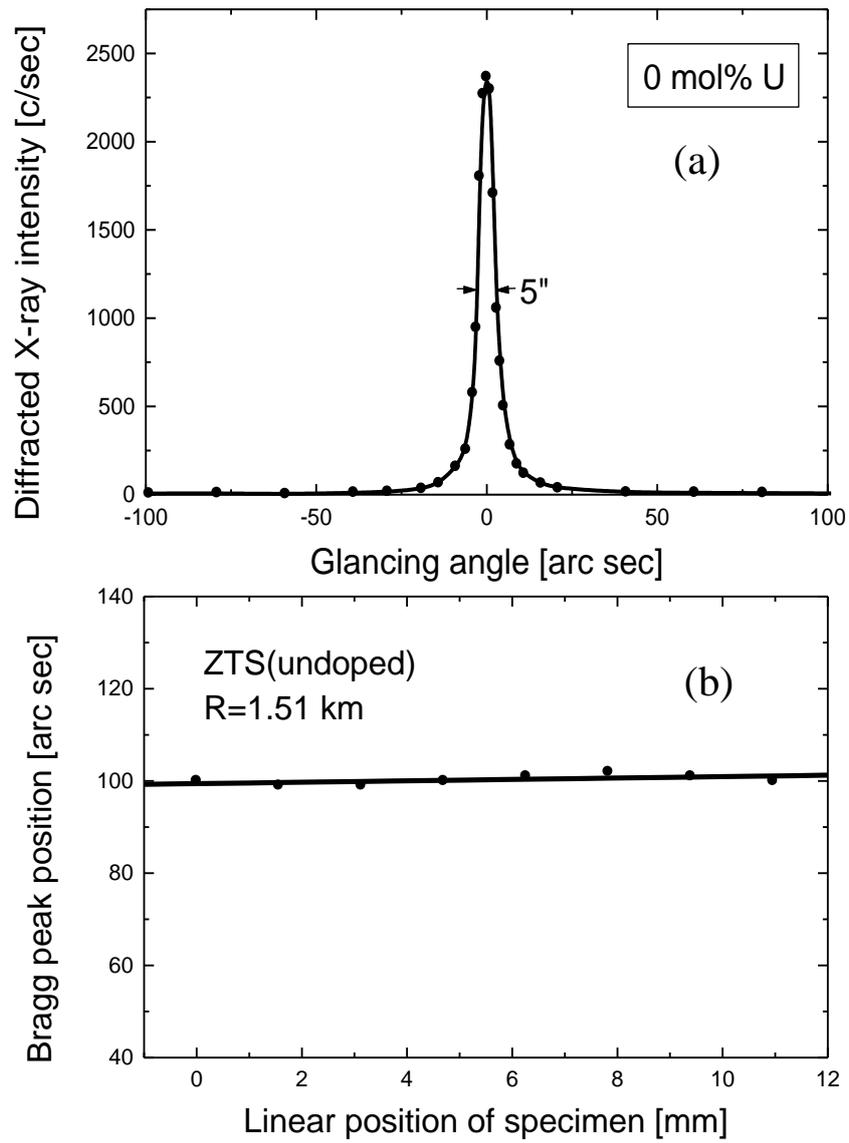

**Figure 3** (a) Diffraction curve recorded for (200) diffracting planes and (b) curvature plot of pure ZTS single crystal.

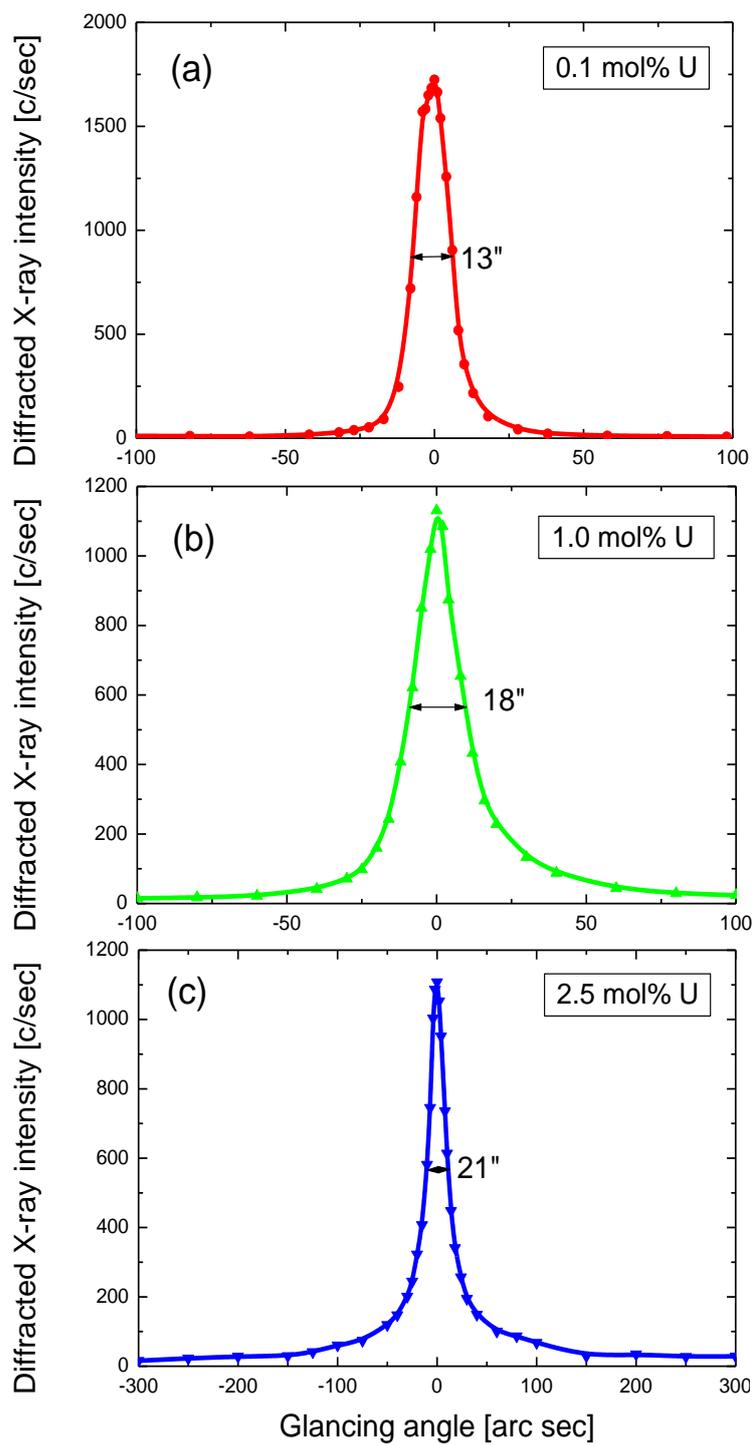

**Figure 4** Diffraction curves of doped ZTS single crystals recorded for (200) diffraction planes: (a) 0.1, (b) 1.0 and (c) 2.5 mol% urea.

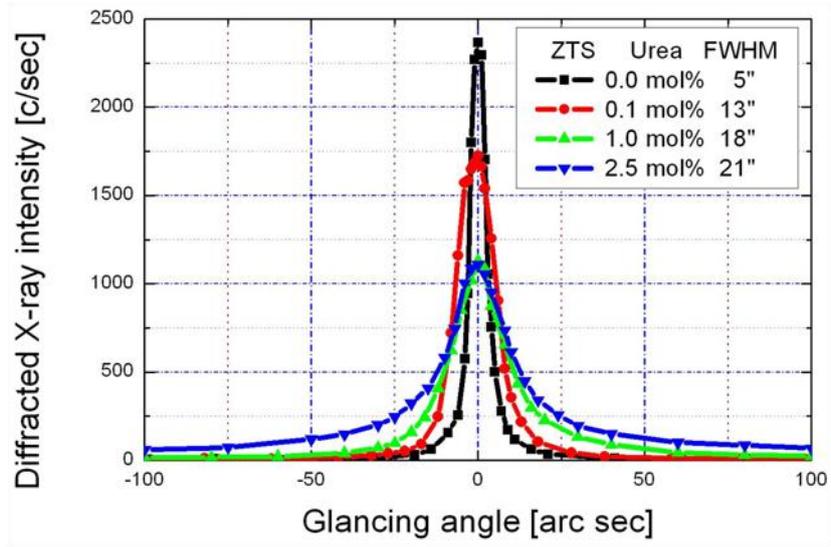

**Figure 5** Comparative representation of diffraction curves for undoped and doped (up to 2.5 mol%) specimens of ZTS crystals using (200) diffracting planes.

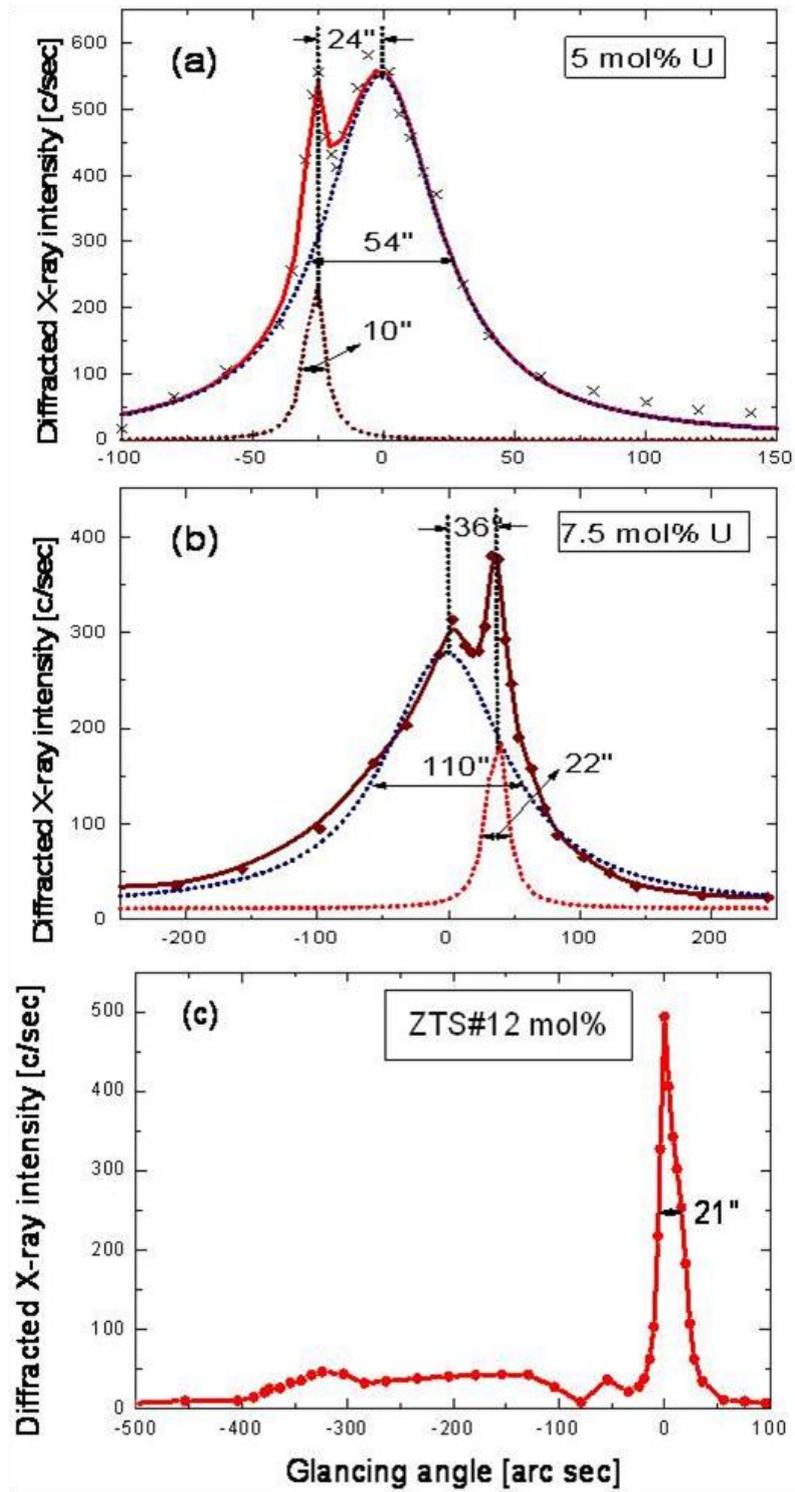

**Figure 6** Diffraction curves (a), (b) and (c) are respectively for 5.0, 7.5 and 12 mol% urea (U) doped ZTS single crystals.

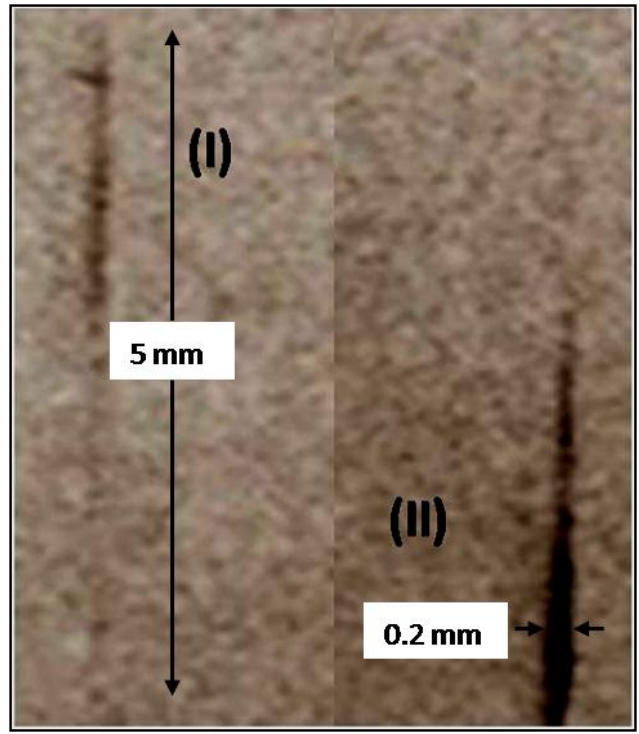

**Figure 7**  Section topographs recorded at the peak positions of the DC of Figure 6(a). The topographs at I and II respectively correspond to the very low angle grain boundary (at 24 arc s away from main peak) and the main peak (at zero position).